\documentclass[twocolumn,showpacs,preprintnumbers,superscriptaddress,amsmath,amssymb]{revtex4}

\usepackage{graphicx}
\usepackage{dcolumn}
\usepackage{bm}
\usepackage{multirow}
\usepackage[none]{hyphenat}

\begin{document}

\preprint{APS/123-QED}

\title{Odd $p$ isotope $^{113}$In: Measurement of Alpha-Induced Reactions}

\author{C. Yal\c c\i n}
\affiliation{Kocaeli University, Department of Physics, Umuttepe
41380, Kocaeli, Turkey\\}%
\affiliation{ATOMKI, H-4001 Debrecen, POB.51., Hungary\\}%

\author{R. T. G\"{u}ray\footnote{Corresponding Author : tguray@kocaeli.edu.tr }}%
\author{N. \"{O}zkan}
\author{S. Kutlu}%
\affiliation{Kocaeli University, Department of Physics, Umuttepe
41380, Kocaeli, Turkey\\}%
\author{Gy. Gy\"{u}rky}
\author{J. Farkas}
\author{G. G. Kiss}
\author{Zs. F\"{u}l\"{o}p}
\author{A. Simon}
\author{E. Somorjai}

\affiliation{ATOMKI, H-4001 Debrecen, POB.51., Hungary\\}%
\author{T. Rauscher}
\affiliation{University of Basel, Department of Physics, CH-4056 Basel, Switzerland\\}%

\date{\today}
             \begin{abstract}
One of the few $p$ nuclei with an odd number of protons is $^{113}$In.
Reaction cross sections of $^{113}$In($\alpha,\gamma)^{117}$Sb and
$^{113}$In($\alpha,n)^{116}$Sb have been
measured with the activation method at center of mass energies
between 8.66 MeV and 13.64 MeV, close to the astrophysically
relevant energy range. The experiments were carried out at
the cyclotron accelerator of ATOMKI. The activities were determined
by off-line detection of the decay gamma rays with a HPGe detector.
Measured cross sections and astrophysical $S$ factor results
are presented and compared with statistical model calculations
using three different $\alpha$+nucleus potentials. The comparison
indicates that the standard rates used in the majority of network
calculations for these reactions were too fast due to the energy
dependence of the optical $\alpha$ potential at low energy.
\end{abstract}


\maketitle

\section{Introduction}
\label{sec:intro}

Astrophysical $s$, $r$, and $p$ processes are thought to synthesize
all nuclei heavier than the iron group. The $s$ (slow
neutron capture) and $r$ (rapid neutron capture) processes proceed
via neutron capture reactions followed by  \mbox{$\beta^{-}$-decays} until
the appearance of stable nuclei. While the $s$ process is responsible for
the production of nuclei along the main $\beta$-stability line only,
the $r$ process contributes to nuclear abundances on the main stability line as well as
its neutron-rich side, not accessible to slow neutron captures.

Additionally, there are 35 rare nuclei along the proton-rich side of the stability
line between Se and Hg, the so-called $p$ nuclei which cannot be
synthesized by neutron captures. Instead, those nuclei have to be created by an
additional process, tentatively called the $p$ process.
A number of different ways have been suggested (see, e.g., \cite{Woosley,Rayet,schatzletter,dauphas,agpp}
and references therein) to synthesize $p$ nuclei, including combinations of
different processes.
Currently the most favored process is photodisintegration in hot
(explosive or non-explosive) shells of massive stars, also called the
$\gamma$ process \cite{Woosley,Rayet}. This $\gamma$ process is governed
by mostly ($\gamma$,n), ($\gamma$,p) and
($\gamma,\alpha$) reactions on preexisting $s$ and $r$ seed
nuclei in the temperature range between \mbox{2 GK} and \mbox{3 GK}. After an initial production
of proton-rich isotopes by ($\gamma$,n) reactions, the synthesis path
branches to nuclides with lower charge $Z$ at isotopes for which ($\gamma$,p) or ($\gamma$,$\alpha$)
reactions are faster than the neutron release.
The ($\gamma$,p) and (p,$\gamma$) reactions are important for the production of the
lower mass $p$ nuclei while ($\gamma$,$\alpha$) reactions contribute to the abundances of medium and heavy mass $p$ nuclei
\cite{Woosley,rappp,rau06,Mohr07}.

The modeling of $p$ (or $\gamma$) process nucleosynthesis requires a large network
of thousands of nuclear reactions involving stable and unstable
nuclei. The relevant astrophysical reaction rates derived
from the reaction cross sections are necessary inputs to this
network. Unfortunately, experimental data for charged-particle
induced reactions are scarce above iron. So far, while more proton
capture reaction cross sections were studied, only a limited number of
$\alpha$-capture reaction cross sections, mostly in the lower mass region,
are available \cite{Basunia05,Fulop96,Rapp02,Gyu06,Ozkan07,Rapp08,Haris05,Danil08,Som98}. The $p$ process
studies are therefore based mostly on Hauser-Feshbach statistical
models to predict the reaction rates. Although the (p,$\gamma$)
measurements generally agree with the statistical model predictions
within less than a factor of two, the ($\alpha,\gamma$)
measurements are considerably lower compared to their
model predictions and indicate that the measured
alpha capture reaction cross sections are not correctly described by global
parameterizations.

\begin{table*}
\caption{\label{tab:decaypar} Decay parameters of the
$^{113}$In\,+\,$\alpha$ reaction products taken from the literature
\cite{nudat, Blachot02} and determined detection efficiency. Only the $\gamma$-transitions used for the analysis are listed.}
\setlength{\extrarowheight}{0.1cm}
\begin{ruledtabular}
\begin{tabular}{ccccc}
\parbox[t]{3.0cm}{\centering{Reaction}} &
\parbox[t]{3.0cm}{\centering{Half-life (minute) }} &
\parbox[t]{3.0cm}{\centering{E$_{\gamma}$ (keV)}} &
\parbox[t]{3.5cm}{\centering{$\gamma$ Emission Prob. ($\%$)}}  &
\parbox[t]{3.0cm}{\centering{Detection Eff. ($\%$)}}  \\

\hline $^{113}$In($\alpha,\gamma)^{117}$Sb & 168.0 $\pm$ 0.6 & 158.562  & 85.9 $\pm$ 0.4 & 1.26 $\pm$ 0.1  \\
$^{113}$In($\alpha$,n)$^{116g}$Sb & 15.8 $\pm$ 0.8 & 931.84 & 24.8 $\pm$ 1.9 & 0.29 $\pm$ 0.02 \\
$^{113}$In($\alpha$,n)$^{116m}$Sb & 60.3 $\pm$ 0.6 & 407.351  & 38.8 $\pm$ 1.6 & 0.58 $\pm$ 0.05 \\
\end{tabular}
\end{ruledtabular}
\end{table*}

The $^{113}$In $p$ nucleus is one of only four $p$ nuclei with an odd number of protons and one of only two with an odd mass number. So far, reactions relevant to the $p$ process have only been investigated with even Z nuclei. Comparisons to Hauser-Feshbach calculations at $p$ process energies have also only been performed for even Z nuclei. Therefore, only reactions with both target and projectile having spin and parity of $J^\pi=0^+$ have been tested. This is the first time an odd-Z target with a non-zero ground state spin has been used.

Furthermore, $^{113}$In has a special importance for the
study of the Cd-In-Sn region. Interpretation of the observed
isotopic abundance in this region in terms of the contributing
nucleosynthesis mechanisms is quite complex due to the
multiple-branched reaction flows in the $s$, $r$, and $p$ process
\cite{rappp,rau06,nemeth94}. Many models show that the initial seed abundance of
$^{113}$In is destroyed by the photodissociation (since the destruction
channel is much stronger than the production channel) and this leads to the
conclusion that $^{113}$In has strong contributions from other processes and even
may not be a $p$ nucleus \cite{Dillmann08}.
There are many studies regarding the nucleosynthesis of $p$ nuclei. The results
concerning $^{113}$In are controversial. While it was synthesized in
sufficient quantities in the model of \cite{Babishov06}, it was
underproduced in other models \cite{Woosley,Rayet,Rayet95,Rausher02}.
In order to understand whether these inconsistencies are
due to nuclear physics inputs or also problems with
astrophysical models, more precisely measured cross sections in the
relevant energy range are necessary.
In this respect, the measurement of $^{113}$In cross sections
also helps to directly understand the problem of the contribution
of the $p$ process in the production or destruction of $^{113}$In in massive stars.

In order to extend the experimental database for the astrophysical
$p$ process and to test the reliability of statistical model
predictions in this mass range, the alpha capture cross sections of
$^{113}$In have been measured in a center of mass energy range
between 8.66 MeV and \mbox{13.64 MeV} using the activation method. These energies
are close to the astrophysically
relevant energy range (the Gamow window) which extends
from \mbox{5.24 MeV} at 2 GK to \mbox{10.17 MeV} at 3 GK.
Reaction rate predictions are very sensitive to the optical model parameters and
this introduces a large uncertainty into theoretical rates involving $\alpha$ particles
at low energy. Therefore, we also compare our new results with
Hauser-Feshbach statistical model calculations using
different $\alpha$+nucleus potentials.

Additionally, $^{113}$In($\alpha,n)^{116}$Sb reaction cross sections were measured. The ($\alpha$,n) cross sections are mainly sensitive to the $\alpha$ width whereas the ($\alpha$,$\gamma$) cross section show a more complicated
dependence on both $\alpha$ and $\gamma$ widths at low energy. A detailed comparison of combined ($\alpha$,$\gamma$) and ($\alpha$,n) data with statistical model calculations allows to better determine the source of possible disagreement between theory and experiment.

Details of our experiment are given in Sec.\ \ref{sec:experiment}. The final
results are presented in Sec.\ \ref{sec:expresults}. A comparison to statistical model
calculations and a detailed discussion is given in Sec.\ \ref{sec:theory}. The final Sec.\ \ref{sec:summary}
provides conclusions and a summary.

\section{Experiment}
\label{sec:experiment}

Since reaction products are radioactive and their half lives are
relatively long, the activation method was used to determine the
reaction cross sections. Experimental aspects for the measurements
of $p$ process reactions by activation methods are discussed in Ref.
\cite{Ozkan02}. The number of $\beta$-unstable isotopes $N_{I}$
produced after each target irradiation for a time period of $t_{I}$
can be obtained by
\begin{equation}
 N_{I} =\displaystyle\sum_{i=1}^n \frac{\phi_{i}\sigma n_{T}} {\lambda} [1-e^{-\lambda \Delta t}]e^{-\lambda(t_{I}-t_{i})}
\end{equation}
where, $\phi_{i}$ is the number of $\alpha$ particles per second
bombarding the target for the time segment $i$, $\Delta$$t$ is the
time interval which is constant for each segment $i$, $\sigma$ is
the reaction cross section, $n_{T}$ is the areal number density of the
target nuclei, $\lambda$ is the decay constant of the product, $n$ is
the total number of time segments in the irradiation period, and
$t_{i}$ is the time length between the beginning of the irradiation and
the end of the $i^{th}$ segment. If the target is counted between
the time $t_{1}$ and $t_{2}$ after the irradiation, the total number
of decays, $N_{D}$:
\begin{equation}
 N_{D} = N_{I}(e^{-\lambda t_{1}} - e^{-\lambda t_{2}})
\end{equation}
Using the decay parameters including the emission probability and the
detection efficiency of an appropriate $\gamma$-transition, the cross
section of the reaction can be determined.

In the case of $^{113}$In($\alpha$,n), the
reaction product $^{116}$Sb has ground and isomeric states. The
partial cross sections leading to these states can be determined
separately because of the different decay patterns of the isomeric
and ground states. The decay parameters used for the analysis are
summarized in Table \ref{tab:decaypar}.

\subsection{Target preparation}
The target were produced by evaporating \mbox{93.10 $\%$} isotopically enriched metallic $^{113}$In (obtained from the company ISOFLEX USA, Certificate No: 49-02-113-1312) onto high purity thin \mbox{(d = 2.4 $\mu$m)} aluminum foils. The In metal piece was evaporated from a carbon crucible heated by DC current. The Al foil was placed \mbox{5.4 cm} above the crucible in a holder defining a circular spot with a diameter of \mbox{12 mm} on the foil for In deposition. The weight of the Al foil was measured before and after the evaporation with a precision better than \mbox{5 $\mu$g}, and then from the difference the \mbox{$^{113}$}In number density could be determined. Altogether five enriched targets were prepared with thicknesses varying between \mbox{168 $\mu$g/cm$^{2}$} and \mbox{289 $\mu$g/cm$^{2}$},
corresponding to the number of \mbox{$^{113}$In} atoms per cm$^{2}$ between \mbox{8.3 x 10$^{17}$} and \mbox{1.4 x 10$^{18}$} with uncertainties between \mbox{7 $\%$} and \mbox{8 $\%$}, respectively, governed by the mass measurement and target inhomogeneity.

One of the targets was also measured by Rutherford backscattering spectrometry (RBS) at the nuclear microprobe of ATOMKI in order to investigate the target homogeneity.  The RBS spectra were taken with a \mbox{2.0 MeV} He$^{+}$ beam of \mbox{3 x 3 $\mu$$m^{2}$} beam spot size and \mbox{500 x 500 $\mu$$m^{2}$} scanning size. Total layer thickness data were extracted from the spectra and their fits. The target thickness varied between its edge and center within \mbox{8 $\%$}.

Due to the relatively low melting point of In, before the measurement test runs were performed with natural In targets to determine the maximally allowed $\alpha$ beam current. The experimental set-up is given in the following subsection. The current was increased from 200 nA to check the target stability that was defined as the ratio of the number of backscattered particles in the $^{113}$In peak in the $\alpha$ spectrum to the total number of counts on the current integrator within the same time period. If this ratio is constant in time, the target keeps its stability. The overall examination of the target took about 6 hours and the target was exposed to the alpha beam for about 2 hours at 1000nA. When the current was increased from 1000 nA to 1300 nA, the ratio dropped very dramatically. These tests showed that there was no target deterioration up to an $\alpha$ beam current of $\mbox{1000 nA}$. However, the target stability was monitored during the all irradiation processes by using the technique described in Section IIB.

\subsection{Activations}

The $^{113}$In targets were irradiated with an $\alpha$ beam starting from
the beam energy \mbox{E = 9.00 MeV} increasing by about \mbox{0.50 MeV} laboratory (lab) energy steps up to \mbox{E = 14.14 MeV}. Laboratory energies have been converted into the effective center of mass energies (E$_{\rm c.m.}^{eff}$) that correspond to beam energies in the target at which one-half of the yield for the full target thickness is obtained \cite{Rolfs87}, and the measurement results are presented versus the effective center of mass energies (Sec.\ \ref{sec:expresults}).

\begin{figure}
\resizebox{0.48\textwidth}{!}{%
\includegraphics[angle=0]{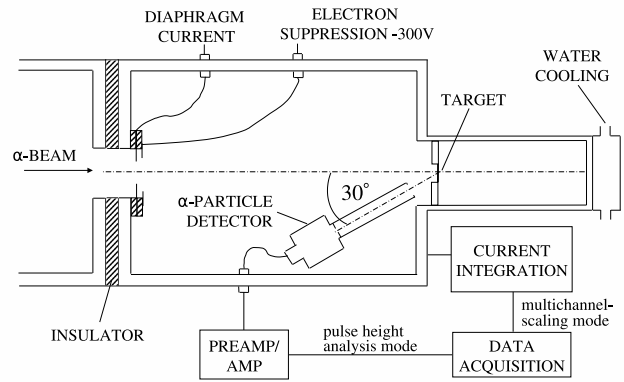}}
\caption{\label{fig:1} A drawing of the target chamber used for the irradiation. }
\end{figure}

A diagram of the target chamber is shown in \mbox{Fig. 1}. After the last beam
defining aperture, the whole chamber was used as a Faraday cup to measure
the beam current. Although the beam current was kept as stable as
possible during the irradiation the beam current was recorded using
a current integrator in order to take into account the possible
changes in the beam current. The integrated current was recorded
every 10 seconds ($\Delta$$t$) by using multi channel scaler. A surface barrier
detector was placed into the target chamber at 150$^{\circ}$ relative to the beam direction to detect backscattered particles and to monitor the target stability. The elastic backscattering spectra were taken continuously and
the number of counts in the $^{113}$In peak in the $\alpha$ spectrum
was checked regularly during the irradiation. There were no substantial background peaks
besides In and Al observed in the spectra. The beam stop was placed
10 cm behind the target from where no backscattered particles could
reach the surface barrier detector. The beam stop was
directly water cooled during the irradiation. A suppression voltage of \mbox{$-300$ V} was applied at the
entrance of the chamber to suppress secondary electrons. Throughout the irradiations with different $\alpha$ beam energies, the typical current was between \mbox{150 nA} and \mbox{800 nA}. The
length of irradiation was chosen based on the longest half-life of the activation products and beam energy, in the range of \mbox{2 h}$-$\mbox{12 h}. Due to steeply decreasing cross
sections at low beam energies, the longer irradiation time was
applied for low-energy measurements to obtain sufficient statistics.

With each target, the first cross section was measured at the lowest energy; the energy was monotonically increased to its maximum value. Since the cross sections increased with energy, this procedure minimized the residual radiation in the target for the subsequent irradiation.  Before usage, each target was checked for activity by counting, to ensure that there was no activity remaining from the previous irradiation.

Since the cyclotron at ATOMKI cannot accelerate the $\alpha$ beam in the beam energy range between about \mbox{10 MeV} and \mbox{11 MeV}, the energy points of \mbox{10.032 MeV}, \mbox{10.565 MeV} and \mbox{11.111 MeV}
were measured with energy degrader foils located \mbox{3 mm} before the target. Aluminum foils with a thickness of \mbox{9.57 $\mu$m} and \mbox{9.70 $\mu$m} were used as energy degrader foils. The thickness of a degrader foil was determined by the RBS technique with a microprobe at the Van de Graaff accelerator of ATOMKI.  The energies \mbox{10.032 MeV}, \mbox{10.565 MeV} and \mbox{11.111 MeV} were reached from the beam energies \mbox{11.000 MeV}, \mbox{11.500 MeV} and \mbox{12.003 MeV}, respectively. The measurements were also made at the beam energies \mbox{9.923 MeV} and \mbox{11.000 MeV} without degrader foils. By comparison to data at energies of \mbox{10.032 MeV} and \mbox{11.111 MeV} obtained \textit{with} degrader foils, the reliability of the degrader method was tested. Both results were found to be very close values at almost the same energies and to show the same energy dependency, as shown in \mbox{Figs. \ref{fig:sfactag} and \ref{fig:sfactan}}.

\subsection{Gamma counting and analysis}

\begin{figure}
\resizebox{0.5\textwidth}{!}{%
\includegraphics{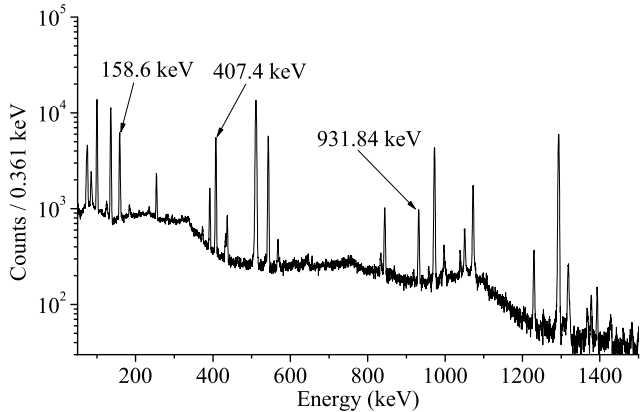}}
\caption{\label{fig:2} Activation $\gamma$  spectrum taken after
irradiating a target with the  $\alpha$ beam of 12 MeV. The
$\gamma$-lines listed in Table I are indicated by arrows. The other
peaks are from either laboratory background or the other
$\gamma$-transitions. \mbox{(1 Channel=0.361keV)}}
\end{figure}

After each irradiation, the target was
taken from the chamber and placed into a low-background counting
area to measure $^{117}$Sb and $^{116}$Sb activities, which are
produced through the $^{113}$In($\alpha,\gamma)^{117}$Sb and
$^{113}$In($\alpha$,n)$^{116}$Sb reactions. The target was placed at
\mbox{3.5 cm} from the end cap of a HPGe detector having \mbox{40 $\%$} relative
efficiency. To reduce the room background, the detector was shielded
with \mbox{10 cm} thick lead bricks. As an example, \mbox{Fig. 2} shows an
off-line  \mbox{$\gamma$-ray} spectrum taken after a \mbox{2.88 h} irradiation
with $\alpha$  beam of \mbox{12 MeV} for a counting time of \mbox{1.24 h}
indicating the  \mbox{$\gamma$-lines} (Table I) used for cross section measurements. For the $^{113}$In($\alpha,\gamma) ^{117}$Sb reaction the \mbox{158.6 keV} $\gamma$-line with the \mbox{85.9 $\%$} emission probability is the only one that has emission probability larger than \mbox{0.3 $\%$} among the others. In the case of $^{113}$In($\alpha,n) ^{116}$Sb reaction, the \mbox{931.8 keV} and \mbox{407.4 keV} $\gamma$-lines were chosen since these two $\gamma$-transitions are associated exclusively
with the decay of ground and isomeric states, respectively.

Absolute efficiency calibration of the detection system was determined at \mbox{10 cm} detector-target distance at which the coincidence-summing effect is negligible. The efficiency-ratio method was used to obtain the efficiencies with
calibrated $^{60}$Co and $^{137}$Cs, uncalibrated $^{152}$Eu and
$^{56}$Co sources. This method requires
a knowledge of the relative emission probabilities of the
multienergy source of unknown activity and at least one energy to
be in an energy range for which the absolute efficiency has
already been determined \cite{Gyu06,Ozkan07,Debertin}. The set of relative
efficiency values (relative to \mbox{122 keV} $\gamma$-efficiency obtained with $^{152}$Eu source and \mbox{847 keV} $\gamma$-efficiency obtained with $^{56}$Co source) was normalized to fit in with
known efficiency values obtained with $^{60}$Co and $^{137}$Cs sources.
In order to normalize the counts for the measurements at the \mbox{3.5 cm} distance to the counts at \mbox{10 cm} distance, an extra irradiation was made at \mbox{14.142 MeV} lab energy, and then the target was counted at both 10 and \mbox{3.5 cm} for the same time period. A factor, which includes geometrical and coincidence-summing ones, was found taking the ratio of the count at \mbox{10 cm} to one at \mbox{3.5 cm} for each $\gamma$-line used for the analysis. Multiplying all measurements at \mbox{3.5 cm} distance by this factor, the detection efficiency at \mbox{10 cm} distance can be used and the coincidence-summing effect is eliminated. The detection efficiencies and the decay properties of the reaction products used for the data analysis are listed in Table \ref{tab:decaypar}. The energy calibration of the detector was done using the efficiency-calibration sources.

\section{Results and Discussion}

\subsection{Measured cross sections and $S$ factors}
\label{sec:expresults}

The $^{113}$In($\alpha,\gamma)^{117}$Sb and
$^{113}$In($\alpha$,n)$^{116}$Sb reaction cross sections have been
measured in the effective center of mass energies between \mbox{8.66 MeV}
and \mbox{13.64 MeV}, which include a part of the astrophysically relevant
energy range. The corresponding astrophysical $S$ factors have also been obtained from the measured cross sections.
The astrophysical $S$ factor is useful for the analysis of charged-particle reactions because it removes part of the strong
energy dependence by accounting for the s-wave Coulomb barrier transmission $\exp (-2\pi \eta)$ at low energies, with $\eta$ being the Sommerfeld parameter. The $S$ factor is defined as \cite{Iliad07}
\begin{equation}
 S(E)=\sigma (E) E e^{2\pi \eta} \quad.
\end{equation}
The experimental results for $^{113}$In($\alpha,\gamma)^{117}$Sb and
$^{113}$In($\alpha$,n)$^{116}$Sb are presented in Tables II and III, respectively. These results provide data for the astrophysical $p$ process network calculations and for a test of statistical models.

\begin{table}
\caption{Measured cross sections and $S$ factors of the
$^{113}$In($\alpha,\gamma)^{117}$Sb reaction.}
\begin{ruledtabular}
\setlength{\extrarowheight}{0.1cm}
\begin{tabular}{lr@{\hspace{0.15cm}}lr@{\hspace{0.15cm}$\pm$\hspace{-0.25cm}}
lr@{\hspace{0.15cm}$\pm$\hspace{-0.25cm}}l} E$_{\rm beam}$ &
\multicolumn{2}{c}{\hspace{-0.2cm}E$_{\rm c.m.}^{eff}$} &
\multicolumn{2}{c}{\hspace{-0.4cm}Cross section} &
\multicolumn{2}{c}{\hspace{-0.5cm}$S$ factor} \\
{[MeV]} & \multicolumn{2}{c}{\hspace{-0.2cm}[MeV]} &
\multicolumn{2}{c}{\hspace{-0.4cm}[$\mu$b]} &
\multicolumn{2}{c}{\hspace{-0.5cm}[$\times$10$^{21}$ MeV b]}
\\ \hline
9.000 & 8.660 & & 3.9 & 0.5 & 402 & 47 \\
9.500 & 9.153 &  & 6.2 & 0.6 & 122 & 11 \\
9.500 & 9.147 &  & 7.4 & 0.6 & 144 &  13 \\
9.923 & 9.553 &  & 15 & 1 & 81 &  7 \\
10.032\footnote{measured with an energy degrader foil.}  & 9.660 &  & 20 & 2 & 74 & 6 \\
10.565\footnotemark[\value {mpfootnote}] & 10.187 & & 37 & 3 & 31 & 3 \\
11.000 & 10.606 & & 64 & 5 & 17 & 1 \\
11.111\footnotemark[\value {mpfootnote}] & 10.704 &  & 70 & 6 & 14 & 1 \\
11.500 & 11.085 & & 111 & 9 & 8.5 & 0.7 \\
12.003 & 11.573 & & 200 & 16 & 4.7 & 0.4 \\
12.612 & 12.162 & & 341 & 28 & 2.2 & 0.2 \\
13.000 & 12.536 & & 435 & 35 & 1.3 & 0.1 \\
13.500 & 13.018 & & 588 & 47 & 0.64 & 0.05 \\
14.142 & 13.640 & & 745 & 60 & 0.25 & 0.02 \\
\end{tabular} \label{tab:ag}
\end{ruledtabular}

\end{table}

\begin{table}
\caption{Measured cross sections and $S$ factors of the
$^{113}$In($\alpha$,n)$^{116}$Sb reaction.}

\begin{ruledtabular}
\setlength{\extrarowheight}{0.1cm}
\begin{tabular}{lr@{\hspace{0.15cm}}lr@{\hspace{0.15cm}$\pm$\hspace{-0.25cm}}
lr@{\hspace{0.15cm}$\pm$\hspace{-0.25cm}}l} E$_{\rm beam}$ &
\multicolumn{2}{c}{\hspace{-0.2cm}E$_{\rm c.m.}^{eff}$} &
\multicolumn{2}{c}{\hspace{-0.4cm}Cross section} &
\multicolumn{2}{c}{\hspace{-0.5cm}$S$ factor} \\
{[MeV]} & \multicolumn{2}{c}{\hspace{-0.2cm}[MeV]} &
\multicolumn{2}{c}{\hspace{-0.4cm}[mb]} &
\multicolumn{2}{c}{\hspace{-0.5cm}[$\times$10$^{21}$ MeV b]}
\\ \hline
10.032\footnote{measured with an energy degrader foil.} & 9.660 &  & 0.07 & 0.02 & 273 & 93 \\
10.565\footnotemark[\value {mpfootnote}] & 10.187 & & 0.27 & 0.04 & 228 & 35 \\
11.000 & 10.606 & & 0.82 & 0.07 & 216 & 21 \\
11.111\footnotemark[\value {mpfootnote}] & 10.704 &  & 1.00 & 0.08 & 199 & 19 \\
11.500 & 11.085 & & 2.1 & 0.2 & 159 & 14 \\
12.003 & 11.573 & & 6.0 & 0.5 & 142 & 14 \\
12.612 & 12.162 & & 13 & 1 & 85 & 8 \\
13.000 & 12.536 & & 25 & 2 & 72 & 7 \\
13.500 & 13.018 & & 50 & 4 & 55 & 5 \\
14.142 & 13.640 & & 88 & 6 & 30 & 3 \\
\end{tabular} \label{tab:ap}
\end{ruledtabular}

\end{table}

The uncertainty in the measurements is based on the following
partial errors: counting statistics (between \mbox{0.2 $\%$} and \mbox{38.9 $\%$}),
detection efficiency ($\thicksim$ 8 $\%$), decay parameters (less
than \mbox{5 $\%$)} and target thickness (between \mbox{7 $\%$} and \mbox{8 $\%$}). The
errors of the beam energy is governed by the energy loss in the
targets determined with the SRIM code \cite{SRIM} (between \mbox{0.2 $\%$}
and \mbox{0.8 $\%$}), uncertainties in the energy degrader foils ($\thicksim$ \mbox{1 $\%$} ) and the energy calibration and stability of the cyclotron
($\thicksim$ \mbox{0.5 $\%$} ). The ($\alpha,\gamma$) reaction of
$^{113}$In was carried out at \mbox{9.5 MeV} with two different targets to
check the systematic uncertainties. The cross section results of
both measurements are in a good agreement (Table II).

The ($\alpha$,n) reactions of $^{113}$In populated the ground state
($T_{1/2}$ = 15.8 min) and isomeric state ($T_{1/2}$ = 60.3 min) of
$^{116}$Sb. The total cross section of the
$^{113}$In($\alpha$,n)$^{116}$Sb reaction was determined by taking
the sum of the partial cross sections of
$^{113}$In($\alpha$,n)$^{116g}$Sb and
$^{113}$In($\alpha$,n)$^{116m}$Sb measured independently using
931.84 keV and 407.351 keV $\gamma$-lines, respectively. The cross sections of these two
($\alpha$,n) reactions are also listed separately in the Table IV.

\begin{table}
\caption{Measured cross sections of the $^{113}$In($\alpha$,n) reactions
that produce ground $^{116g}$Sb and isomeric $^{116m}$Sb states. For
the analysis, \mbox{931.84 keV} and \mbox{407.351 keV} $\gamma$-transitions,
respectively, were used: for the decay parameters see Table I.}
\begin{ruledtabular}
\setlength{\extrarowheight}{0.1cm}
\begin{tabular}{lccc}
\multicolumn{1}{c}{E$_{\rm beam}$} &
\multicolumn{1}{c}{E$_{\rm c.m.}^{eff}$} &
\multicolumn{2}{c}{Cross Section (10$^{-6}$ barn)} \\
\multicolumn{1}{c} {(MeV)} &
\multicolumn{1}{c} {(MeV)} &
\multicolumn{1}{c}{$^{116m}$Sb (407.351 keV) } &
\multicolumn{1}{c} {$^{116g}$Sb}(931.84 keV)
\\
\hline
10.032\footnote{measured with an energy degrader foil.} & 9.660 & 17.1 $\pm$ 2.6 & 55 $\pm$ 22 \\
10.565\footnotemark[\value {mpfootnote}] & 10.187 & 68.3 $\pm$ 6.3 & 205 $\pm$ 35 \\
11.000 & 10.606 & 173 $\pm$ 15 & 654 $\pm$ 66 \\
11.111\footnotemark[\value {mpfootnote}] & 10.704 & 213 $\pm$ 17 & 790 $\pm$ 76 \\
11.500 & 11.085 & 459 $\pm$ 37 & 1622 $\pm$ 150 \\
12.003 & 11.573 & 1324 $\pm$ 107 & 4660 $\pm$ 490 \\
12.612 & 12.162 & 3763 $\pm$ 306 & 9657 $\pm$ 1030 \\
13.000 & 12.536 & 6662 $\pm$ 539 & 18197 $\pm$ 1915 \\
13.500 & 13.018 & 14585 $\pm$ 1190 & 35395 $\pm$ 3693 \\
14.142 & 13.640 & 27393 $\pm$ 2255 & 60236 $\pm$ 5720 \\
\end{tabular} \label{tab:an}
\end{ruledtabular}
\end{table}

\subsection{Comparison with Hauser-Feshbach predictions}
\label{sec:theory}

\begin{figure}
\resizebox{0.52\textwidth}{!}{%
\includegraphics {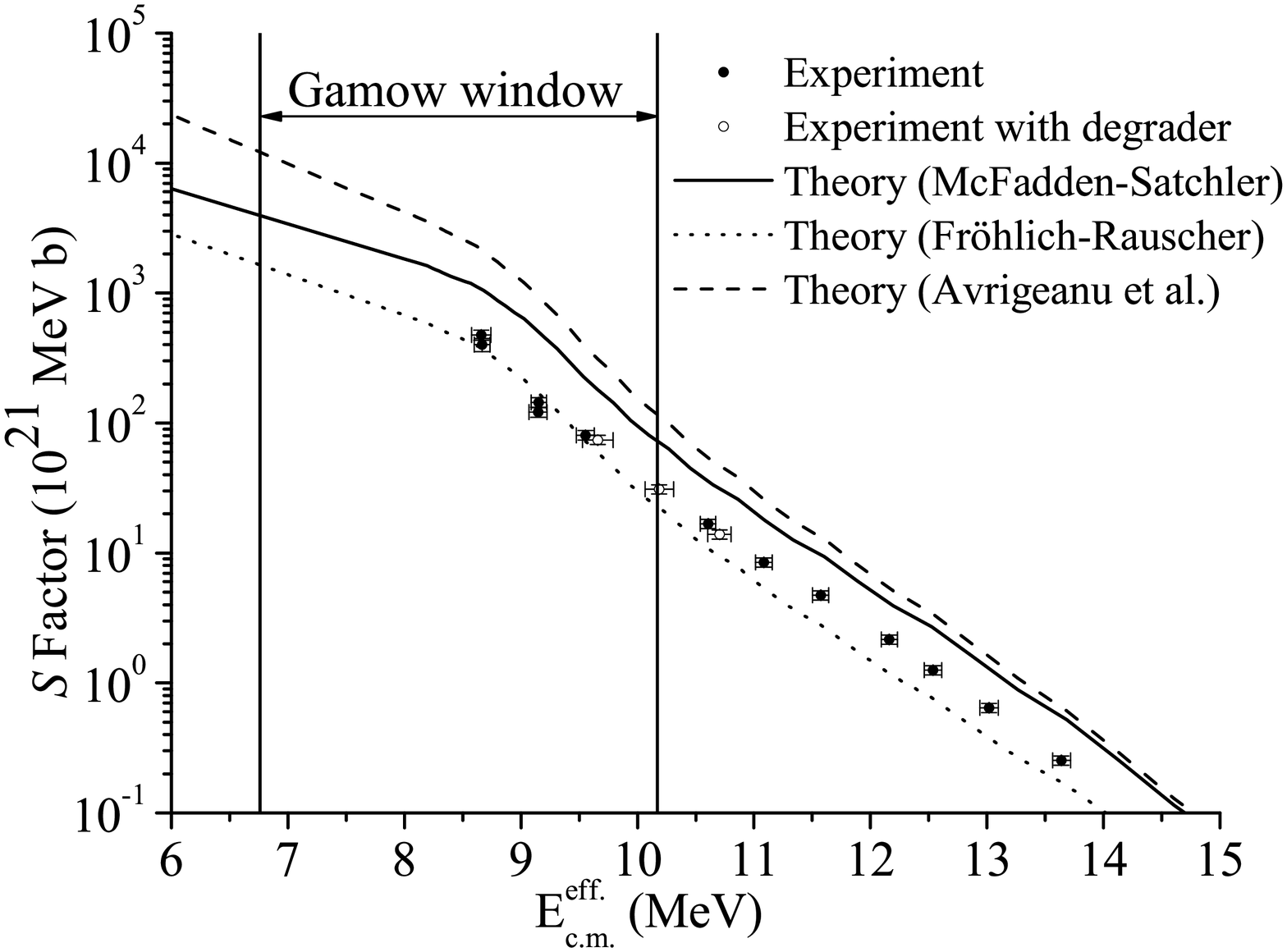}}
\caption{\label{fig:sfactag}Measured $S$ factors of $^{113}$In($\alpha$,$\gamma$)$^{117}$Sb
reaction compared to theory using the NON-SMOKER$^\mathrm{WEB}$ v5.4.2w code
\cite{nonsmokerweb} with different $\alpha$+nucleus potentials: by McFadden
and Satchler \cite{McFadden}, Fr\"{o}hlich \cite{Frohlich,Rauscher03}, and Avrigeanu et al. \cite{Avrigeanu}.
Also shown is the astrophysically relevant energy range (Gamow window) for a stellar temperature of \mbox{3 GK}.}

\end{figure}

\begin{figure}
\resizebox{0.52\textwidth}{!}{%
\includegraphics {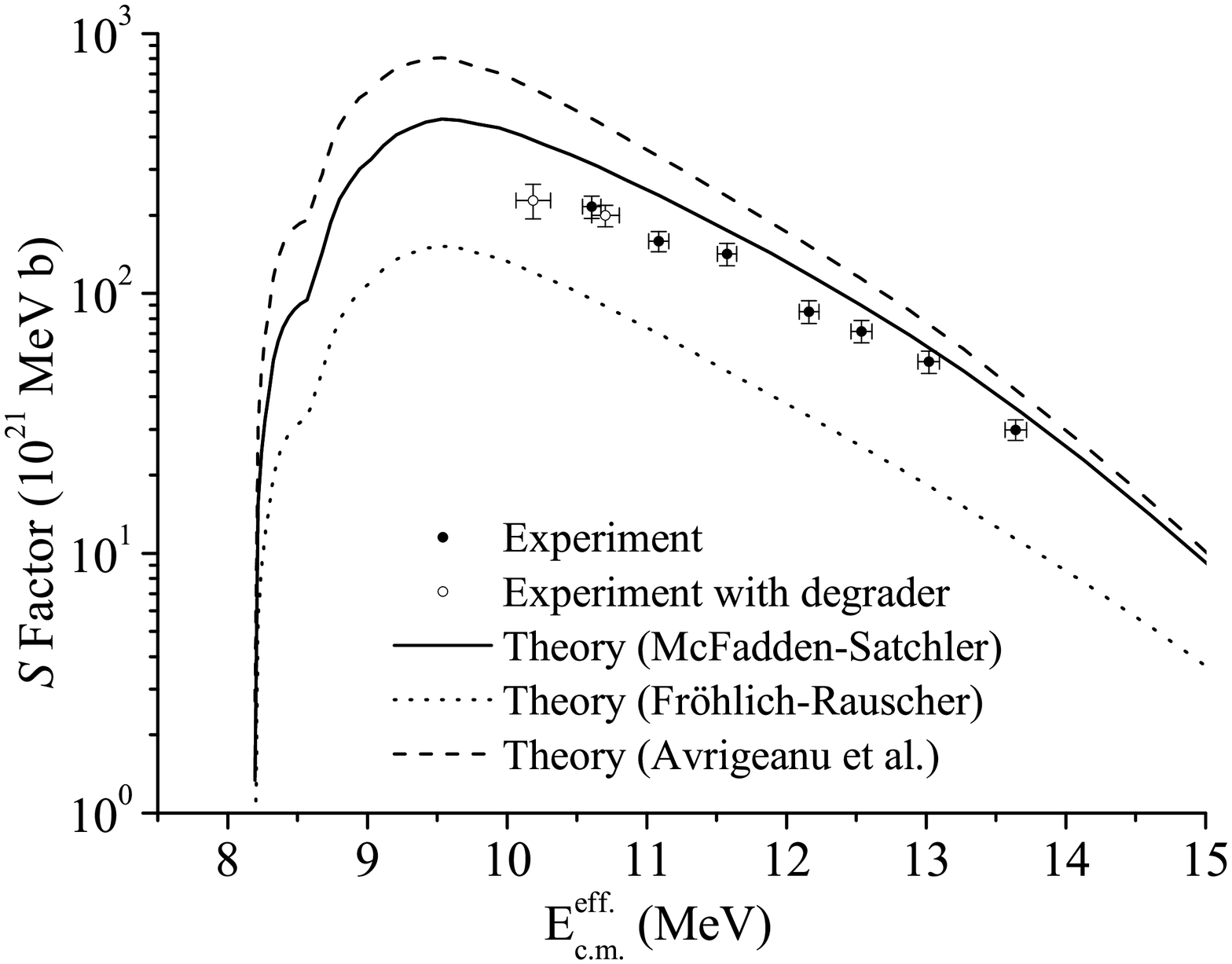}}
\caption{\label{fig:sfactan}Measured $S$ factors of $^{113}$In($\alpha$,n)$^{116}$Sb
reaction compared to theory using the NON-SMOKER$^\mathrm{WEB}$ v5.4.2w code
\cite{nonsmokerweb} with different $\alpha$+nucleus potentials: by McFadden
and Satchler \cite{McFadden}, Fr\"{o}hlich \cite{Frohlich,Rauscher03}, and Avrigeanu et al. \cite{Avrigeanu}. }
\end{figure}

For the $^{113}$In($\alpha,\gamma)^{117}$Sb and
$^{113}$In($\alpha$,n)$^{116}$Sb reactions, the measured
astrophysical $S$ factors have been compared with the
Hauser-Feshbach statistical model calculations obtained with the
statistical model code NON-SMOKER$^\mathrm{WEB}$ \cite{nonsmokerweb} (an update and upgrade of the previous NON-SMOKER code \cite{RT98,Raus00,Raus01}), version v5.4.2w, with different, frequently used $\alpha$+nucleus potentials as seen in
\mbox{Figs.\ \ref{fig:sfactag} and \ref{fig:sfactan}:} McFadden-Satchler \cite{McFadden} (which is the standard setting), Fr\"{o}hlich-Rauscher \cite{Frohlich,Rauscher03}, and Avrigeanu et al. \cite{Avrigeanu}. It has to be noted that both the potentials \cite{McFadden} and \cite{Avrigeanu} have been fitted to scattering data at
higher energies, \cite{McFadden} to data above \mbox{70 MeV} for a wide range of nuclei, \cite{Avrigeanu} to data above \mbox{14 MeV} for nuclei around \mbox{$A\approx 100$}. The potential of \cite{Frohlich,Rauscher03}, on the other hand, has been fitted to reaction data for nuclei in the mass range \mbox{$144\leq A\leq 157$} but was found to reproduce $\alpha$-induced low-energy reaction cross sections well also for targets outside of this range.

\mbox{Figure \ref{fig:sfactag}} shows that the calculations with the potentials of
\cite{McFadden} and \cite{Avrigeanu} overestimate the ($\alpha,\gamma$) $S$ factors
by maximally a factor of 1.8 and 8.3, respectively.
For the potential of \cite{Frohlich}, although an agreement is
observed at lower energies, the theoretical prediction deviates from
the experimental data as the energy increases.

The case of the ($\alpha$,n) reaction is shown in \mbox{Fig.\ \ref{fig:sfactan}}. The
calculations with the potential of \cite{McFadden} have best
agreement with the measured $S$ factors. The predictions with
the potential of \cite{Avrigeanu} are higher than the experimental
results by factors of between 1.5 and 2.1, with closer agreement at higher energy, while the obtained values
with the potential of \cite{Frohlich} underestimate the
measurements by factors from 3.5 to 4.4, with better agreement at lower energy. The energy dependence of the $S$ factor is overall reproduced satisfactorily by the potentials \cite{McFadden} and \cite{Frohlich} when
appropriately scaled. The potential of \cite{Avrigeanu} yields an increase in the $S$ factor which is
too steep when going to low energy. Note, however, that the predictions with the potentials of \cite{McFadden}
and \cite{Avrigeanu} become quite similar above 14 MeV.

\subsection{$\alpha$ and $\gamma$ width sensitivities of the predictions}

Hauser-Feshbach cross sections depend on a number of nuclear properties which have to be known
experimentally or predicted by a model. NON-SMOKER$^\mathrm{WEB}$ is a global code in its standard setting, not fine-tuned to any local
parameters and aiming at the best global description for a wide range of nuclei although there may be some deviations encountered when looking
at individual reactions in an isolated manner. In order to understand the sensitivities and to disentangle the different contributions of
nuclear properties or transitions, it has to be remembered that the central quantities in the statistical model are transmission coefficients $\mathcal{T}$ being related to averaged widths $\mathcal{T}=2\pi \rho \left< \Gamma \right>$, where $\rho$ is a nuclear level density. The cross section (or $S$ factor) is related to the transmission coefficients in the entrance and exit channel \cite{Koehler04,Descouvemont06}
\begin{equation}
\label{eq:haufesh}
 \sigma \propto \frac{\mathcal{T}_\mathrm{entrance} \mathcal{T}_\mathrm{exit}}{\mathcal{T}_\mathrm{tot}} \propto
\frac{\left< \Gamma_\mathrm{entrance} \right>\left< \Gamma_\mathrm{exit} \right>}{\left< \Gamma_\mathrm{tot} \right>} \quad,
\end{equation}
where the subscript ``$\mathrm{tot}$'' labels the total quantities which include all energetically accessible, ``open'' channels.

In our case, the entrance channel is always the system formed by an $\alpha$ particle and the target nucleus whereas the exit channel either
contains a $\gamma$ or a neutron plus the final nucleus, depending on the considered reaction. Particle transmission coefficients are
calculated by solving the Schr\"odinger equation using an optical potential, radiative transmission coefficients include a sum of possible
$\gamma$ transitions with strengths derived from theoretical descriptions (see, e.g., \cite{Raus00,Descouvemont06} for details). In principle, also the nuclear level density enters both types of transmission coefficients via a sum over possible final states. However, the relevant final states are experimentally known for the reactions discussed in this paper and no theoretical nuclear level density has to be invoked. For a given reaction and projectile energy it is not always obvious which (energy dependent) transmission coefficients (or average widths) contribute most to the cross section according to the above equation and thus dominate the sensitivity. Similar considerations apply to Eq.\ (\ref{eq:haufesh}) as to the well-known dependence of Breit-Wigner resonant cross sections \cite{Descouvemont06}. When the widths in the numerator are of very different
size and the larger width dominates the total width, the larger width will cancel out and the cross section depends only on the smaller width. The dependence is more complicated if both widths are of similar size or another open channel significantly contributes to the total width. Since the widths change with energy, also the sensitivity is energy dependent.

\begin{figure}
\resizebox{0.52\textwidth}{!}{%
\includegraphics {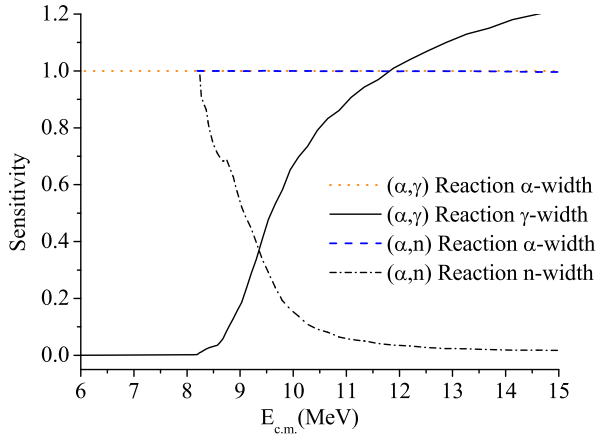}}
\caption{\label{fig:sensi}(Color online) Sensitivity of the astrophysical $S$ factor for the ($\alpha$,$\gamma$) and  ($\alpha$,n) reactions on $^{113}$In to a variation in the $\alpha$, $\gamma$, and neutron widths, as a function of center-of-mass energy.}
\end{figure}

In order to inspect the sensitivity of the
astrophysical $S$ factors for the $^{113}$In($\alpha$,$\gamma$) and $^{113}$In($\alpha$,n) reactions to changes in
the $\alpha$, neutron, and $\gamma$ width, additional calculations were performed, starting from the standard prediction (using the potential of McFadden and Satchler \cite{McFadden}) and independently varying the widths by
factors 0.5 and 2. Here, the sensitivity $\delta$ is defined as the
ratio of relative change in the $S$ factor to one in the width,
\begin{equation}
\delta=\frac{ \Delta S / S } { \Delta \Gamma / \Gamma } \quad,
\end{equation}
where $\Delta S$ is the change in the $S$ factor and  $\Delta \Gamma$ is the change in the width $\Gamma$. \mbox{Figure \ref{fig:sensi}} shows the $\alpha$, neutron, and $\gamma$ width sensitivities of the $S$ factors for the $^{113}$In($\alpha$,$\gamma$) and $^{113}$In($\alpha$,n) reactions: Zero sensitivity means there is no change in the $S$ factor when the width is varied while $\delta =1$ means the $S$ factor changes by the same factor as the width.

\begin{figure}
\resizebox{0.52\textwidth}{!}{%
\includegraphics {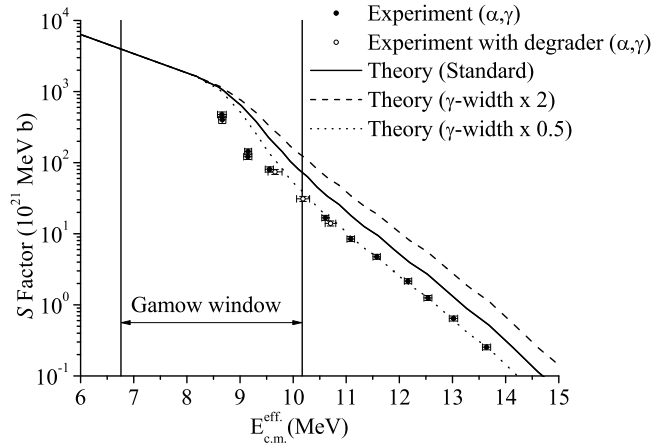}}
\caption{\label{fig:variaggam}$S$ factors for the ($\alpha,\gamma$) reaction obtained using NON-SMOKER$^\mathrm{WEB}$ v5.4.2w
code with two different variations of $\gamma$ width by factors 0.5
and 2, as well as the experimental values. Also shown is the astrophysically relevant energy range (Gamow window) for a stellar temperature of \mbox{3 GK}.}
\end{figure}

\begin{figure}
\resizebox{0.52\textwidth}{!}{%
\includegraphics {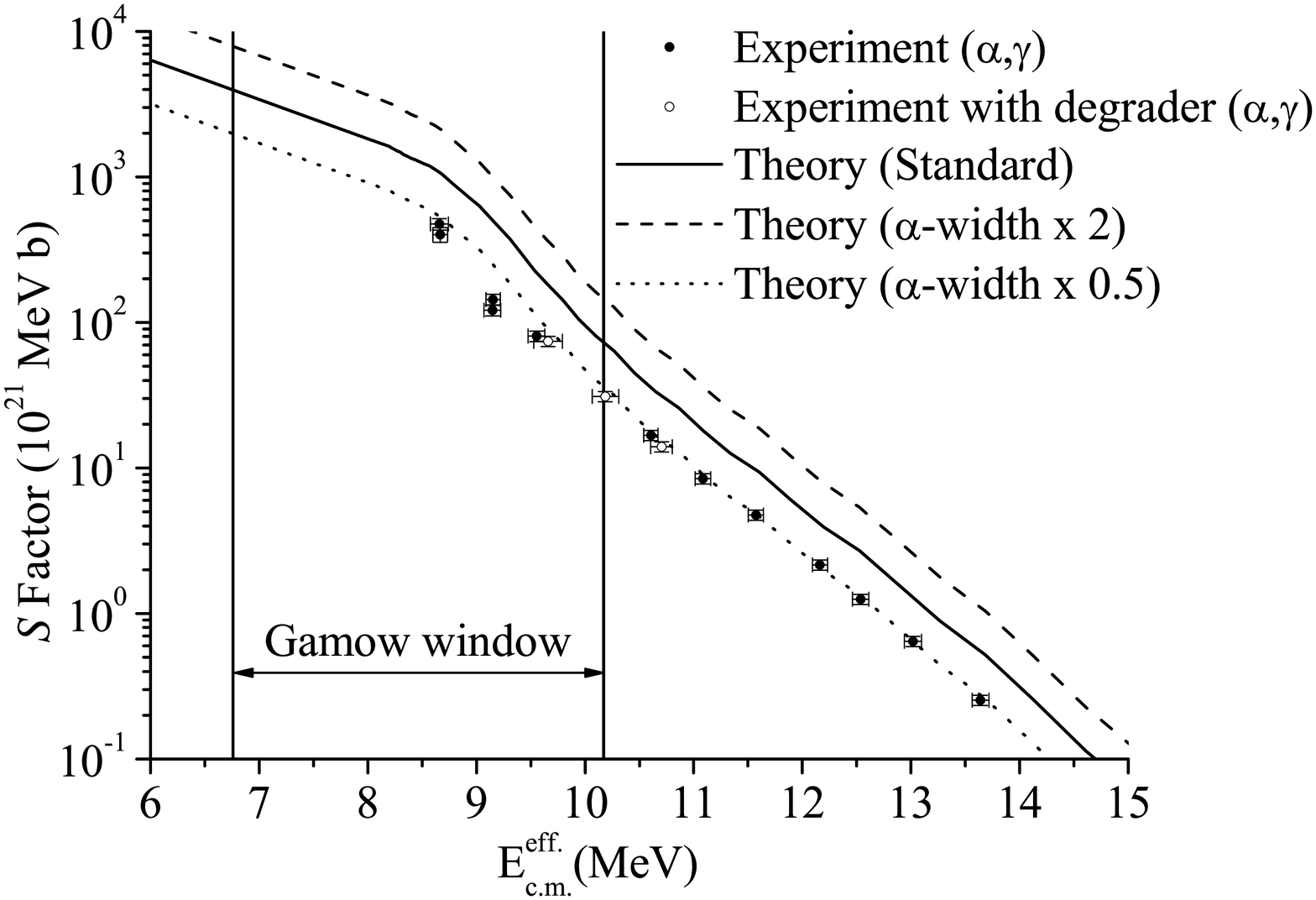}}
\caption{\label{fig:variagalp}$S$ factors for the ($\alpha,\gamma$) reaction obtained using NON-SMOKER$^\mathrm{WEB}$ v5.4.2w
code with two different variations of $\alpha$ width by factors 0.5
and 2, as well as the experimental values. Also shown is the astrophysically relevant energy range (Gamow window) for a stellar temperature of \mbox{3 GK}.}
\end{figure}

\begin{figure}
\resizebox{0.52\textwidth}{!}{%
\includegraphics {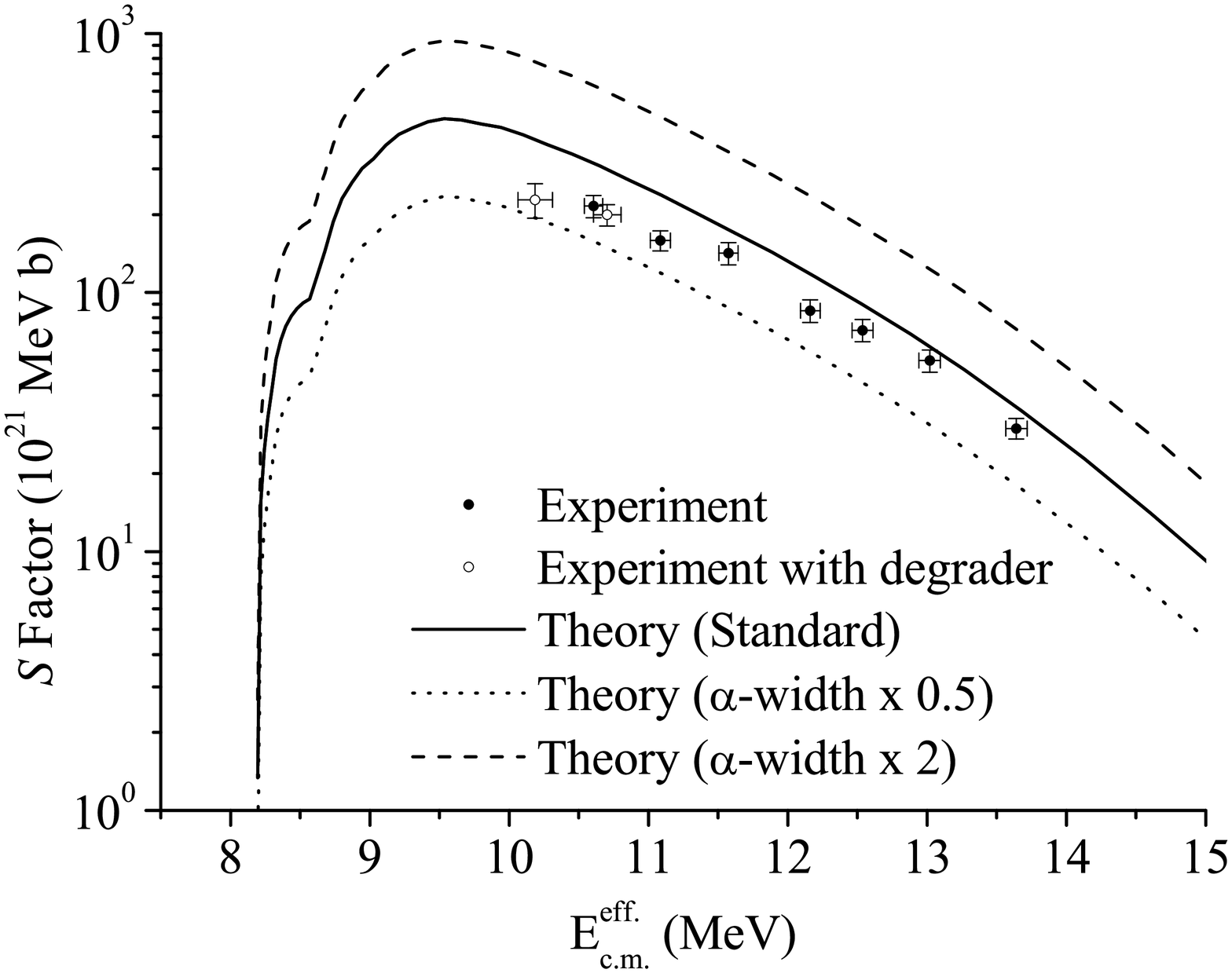}}
\caption{\label{fig:varianalp}$S$ factors for the ($\alpha$,n) reaction obtained using NON-SMOKER$^\mathrm{WEB}$ v5.4.2w
code with two different variations of $\alpha$ width by factors 0.5
and 2, as well as the experimental values.}
\end{figure}

\begin{figure}
\resizebox{0.52\textwidth}{!}{%
\includegraphics {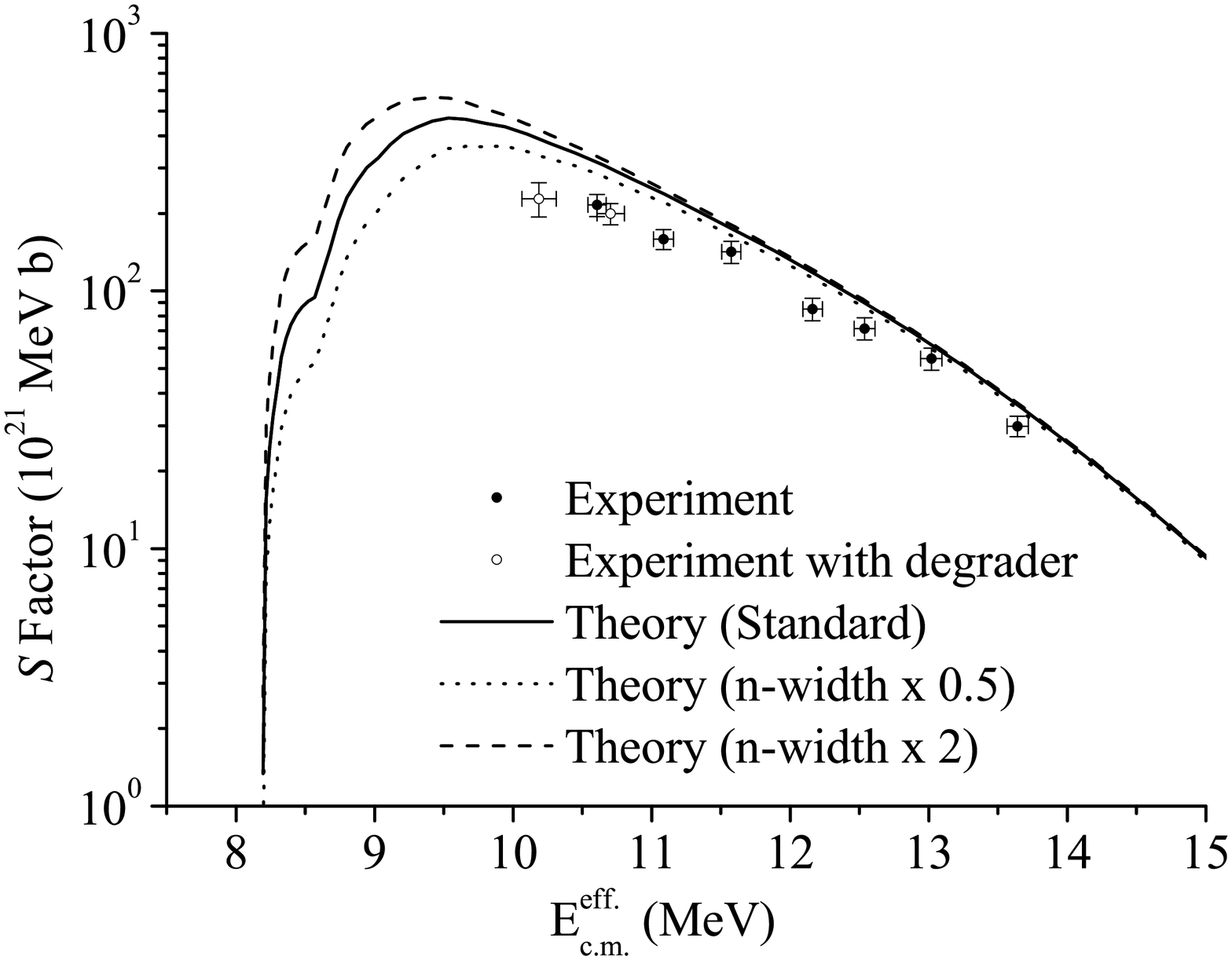}}
\caption{\label{fig:varianneu}$S$ factors for the ($\alpha$,n) reaction obtained using NON-SMOKER$^\mathrm{WEB}$ v5.4.2w
code with two different variations of n width by factors 0.5
and 2, as well as the experimental values.}
\end{figure}

At low energies, the $S$ factor for the $^{113}$In($\alpha$,$\gamma$)$^{117}$Sb reaction is
mainly sensitive to the variation in $\alpha$ width, as seen in \mbox{Fig.\ \ref{fig:sensi}}, since the $\gamma$ width is larger than the $\alpha$ width due to the Coulomb barrier hindering $\alpha$ particles to form a compound nucleus. The Coulomb repulsion is easier overcome at higher energy, leading to an increase in the $\alpha$ width whereas the $\gamma$ width is much less energy dependent. Therefore, the sensitivity to the $\gamma$ width increases with increasing energy. Above the neutron threshold also the neutron width quickly increases and dominates the total width. This leads to a rescaling and keeps the sensitivity to both $\alpha$ and $\gamma$ width high. Therefore, at the lowest energies (and within the
astrophysically relevant Gamow window) the $S$ factor is almost exclusively dependent on the $\alpha$ width while it is sensitive to both $\alpha$
and $\gamma$ width above the Gamow window. This is why a variation of the $\gamma$ width will not change the $S$ factor at low energy, as shown
in \mbox{Fig.\ \ref{fig:variaggam}}, whereas a change in the $\alpha$ width will impact the $S$ factor almost equally at all energies, as seen in \mbox{Fig.\ \ref{fig:variagalp}}.

The sensitivity of the $^{113}$In($\alpha$,n) reaction is less complex than that of the capture reaction. It is also shown in \mbox{Fig.\ \ref{fig:sensi}}. The $S$ factors are sensitive to the $\alpha$ width across the full energy range, similar to the case of the ($\alpha$,$\gamma$) reaction. There is appreciable sensitivity to the
neutron width only in a small energy range above the neutron threshold. Towards higher energy, the neutron width becomes large, will dominate the
total width and thus cancels out. This can also be seen in \mbox{Figs.\ \ref{fig:varianalp} and \ref{fig:varianneu}} which show the absolute changes of the predicted $S$ factors when varying the $\alpha$ and neutron width, respectively.

Pondering the above sensitivities it becomes obvious that the ($\alpha$,n) experiment -- although astrophysically of minor importance -- is useful to test the optical $\alpha$ potential which enters the calculation of the $\alpha$ widths. The capture reaction shows a more complicated dependence
on \textit{three} widths, the $\alpha$, $\gamma$, \textit{and} neutron width. Unfortunately, the $^{113}$In($\alpha$,n) reaction does not allow
to test the optical potential at astrophysically relevant energies because of the neutron threshold at positive energy. Contrary to the first impression one might get from \mbox{Fig.\ \ref{fig:sfactag}}, a combined study of \mbox{Figs.\ \ref{fig:sfactan} and \ref{fig:varianneu}} reveals that the
potential of \cite{McFadden} actually reproduces the data best. Only a slight rescaling is needed to reproduce the $S$ factor of the ($\alpha$,n) reaction. Only the data point at the lowest energy may hint at an incorrect energy dependence towards even lower energies. However, this may also be caused by an incorrect neutron width just above the channel opening.

Taking into account the above conclusion, the measured capture $S$ factor shows that the $\gamma$ width seems to be predicted too large. However, a rescaling of the $\gamma$ width will not affect the $S$ factor at the two lowest measured energies (see \mbox{Fig.\ \ref{fig:variaggam}}). At these two energies, the neutron width is also small and may have an impact. As shown in \mbox{Fig.\ \ref{fig:varianneu}}, to decrease the low-energy $S$ factor of the ($\alpha$,n) reaction and to bring it in accordance with the data, a decrease of the neutron width is needed. However, a smaller neutron width would lead to an even larger $S$ factor at low energy in the ($\alpha$,$\gamma$) reaction because in the capture reaction the neutron width only appears in the denominator of Eq.\ (\ref{eq:haufesh}) as part of the total width. This indicates that the $\alpha$ width is predicted too large at the smallest energies and this can account for the low-energy behavior in both reactions. Interestingly, it appears as if the potential of \cite{Frohlich} provides the required $\alpha$ width at those lowest energies (see \mbox{Fig.\ \ref{fig:sfactag}}) although its energy dependence is not suited to reproduce the ($\alpha$,n) data at slightly higher energy (see \mbox{Fig.\ \ref{fig:sfactan}}). To explain the ($\alpha$,$\gamma$) across the full range of measured energies, a combination of a smaller $\alpha$ width and a smaller $\gamma$ width can reproduce the data.

\section{Summary and Conclusion}
\label{sec:summary}

Using the activation method, the cross sections of the reaction $^{113}$In($\alpha$,$\gamma$)$^{117}$Sb was measured from \mbox{$8.66-13.64$ MeV} effective center-of-mass energy and the ones of the reaction $^{113}$In($\alpha$,n)$^{116}$Sb in the energy range \mbox{$9.66-13.64$ MeV}. This is
at the upper end of the astrophysically relevant energy window. The results were compared to Hauser-Feshbach calculations with different optical $\alpha$+nucleus potentials. A sensitivity study was performed by varying particle and $\gamma$ widths.

We conclude that the combined ($\alpha$,$\gamma$) and ($\alpha$,n) data is acceptably well described by the potential of \cite{McFadden}, except at the lowest energies where we see an indication that the energy dependence is incorrect and leads to overestimated $S$ factors. It appears as if there is a transition from a potential close to the one of \cite{McFadden} to a potential close to the one of \cite{Frohlich} within the Gamow window. We also found that the predicted $\gamma$ width is too large but this is irrelevant at astrophysical energies.

Since the reverse rate varies in the same manner as the forward rate according to the principle of detailed balance, the photodisintegration rate will also become lower when the low-energy ($\alpha$,$\gamma$) $S$ factors become smaller \cite{Raus00}. Although $^{113}$In can be produced by $^{117}$Sb($\gamma$,$\alpha$)$^{113}$In, as studied here, the direct impact of a changed rate on $p$ process calculations will be limited because the $^{117}$Sb($\gamma$,n)$^{116}$Sb channel is considerably faster \cite{rau06}. A similar situation occurs for the destruction of $^{113}$In by ($\gamma$,$\alpha$) and ($\gamma$,n) reactions. Nevertheless, problems with the optical alpha potential and the prediction of low-energy rates similar to the ones found here are expected for nuclei comprising ($\gamma$,$\alpha$) branchings in the $p$ process path. Further measurements to lower energy with $^{113}$In and other targets are required to globally determine the actual energy dependence of the optical $\alpha$ potential at low energy and thus improve the prediction of astrophysical $S$ factors and reaction rates.
\\
\\

\begin{acknowledgments}

The authors thank Jim Fitzgerald for giving permission to use
FitzPeaks Gamma Analysis Software.
This work was supported by The Scientific and Technological Research
Council of Turkey TUBITAK [Grant-108T508 and grant
TBAG-U/111(104T2467)], Kocaeli University BAP (grant 2007/37, 2007/36),
Erasmus (LLLP), the European Research Council grant agreement no.
203175, the Economic Competitiveness Operative Programme
GVOP-3.2.1.-2004-04-0402/3.0, OTKA (K68801, T49245) and the Swiss
NSF (grant 2000-105328). Gy.\ Gy.\ acknowledges support from the Bolyai
grant.
\end{acknowledgments}


\begin{thebibliography}{}
%
\bibitem{Woosley}
S. E. Woosley and W. M. Howard, Astrophys.\ J. Suppl.\ Ser.\ \textbf{36}, 285
(1978).

\bibitem{Rayet}
M. Rayet, N. Prantzos, and M. Arnould, Astron.\ Astrophys.\
\textbf{227}, 271 (1990).

\bibitem{schatzletter} H. Schatz, A. Aprahamian, V. Barnard, L. Bildsten, A. Cumming, M. Ouellette,
T. Rauscher, F.-K. Thielemann, and M. Wiescher, Phys.\ Rev.\ Lett.\ \textbf{86}, 3471 (2001).

\bibitem{dauphas} N. Dauphas, T. Rauscher, B. Marty, and L. Reisberg, Nucl.\ Phys.\ \textbf{A719}, 287 (2003).

\bibitem{agpp} M. Arnould and S. Goriely, Phys.\ Rep.\ \textbf{384}, 1 (2003).

\bibitem{rappp} W. Rapp, J. G\"orres, M. Wiescher, H. Schatz, and F. K\"appeler, Astrophys.\ J. \textbf{653}, 474 (2006).

\bibitem{rau06} T. Rauscher, Phys.\ Rev.\ C \textbf{73}, 015804 (2006).

\bibitem{Mohr07}
P. Mohr, Zs. F\"{u}l\"{o}p, H. Utsunomiya, Eur.\ Phys.\ J. A
\textbf{32}, 357 (2007).

\bibitem{Basunia05}
M.S. Basunia, E. B. Norman, H. A. Shugart, A. R. Smith, M. J.
Dolinski, and B. J. Quiter, Phys.\ Rev.\ C \textbf{71}, 035801 (2005).

\bibitem{Fulop96}
Zs. F\"{u}l\"{o}p, A. Z. Kiss, E. Somorjai, C. E. Rolfs, H. P.
Trautvetter, T. Rauscher, and H. Oberhummer, Z. Phys.\ A
\textbf{355}, 203 (1996).

\bibitem{Rapp02}
W. Rapp, M. Heil, D. Hentschel,  F. K\"{a}ppeler, R. Reifarth, H. J.
Brede, H. Klein, and T. Rauscher, Phys. Rev. C \textbf{66}, 015803
(2002).

\bibitem{Gyu06}
Gy. Gy\"{u}rky, G. G. Kiss, Z. Elekes, Zs. F\"{u}l\"{o}p, E.
Somorjai, A. Palumbo, J. G\"{o}rres, H. Y. Lee, W. Rapp, M.
Wiescher, N. \"{O}zkan, R.T. G\"{u}ray, G. Efe, and T. Rauscher,
Phys. Rev. C \textbf{74}, 025805 (2006).

\bibitem{Rapp08}
W. Rapp, I. Dillmann, F. K\"{a}ppeler, U. Giesen, H Klein, T.
Rauscher, D. Hentschel and S. Hilpp, Phys. Rev. C \textbf{78},
025804 (2008).

\bibitem{Ozkan07}
N. \"{O}zkan, G. Efe, R.T. G\"{u}ray, A. Palumbo, J. G\"{o}rres, H.
-Y. Lee, L. O. Lamm, W. Rapp, E. Stech, M. Wiescher, G. Gy\"{u}rky, Zs.
F\"{u}l\"{o}p, E. Somorjai, Phys. Rev. C \textbf{75}, 025801 (2007).

\bibitem{Haris05}
S. Harissopulos, A. Lagoyannis, A. Spyrou, Ch. Zarkadas, G.
Galanopoulos, G. Perdikakis, H-W Becker, C. Rolfs et al., J. Phys.
G.: Nucl. Part. Phys. \textbf{31}, S1417 (2005).

\bibitem{Danil08}
I. Cata-Danil, D. Filipescu, M. Ivascu, D. Bucurescu, N. V. Zamfir,
T. Glodariu, L. Stroe, G. Cata-Danil, D. G. Ghita, C. Mihai, G.
Suliman, and T. Sava, Phys. Rev. C \textbf{78}, 035803 (2008).

\bibitem{Som98}
E. Somorjai, Zs. F\"{u}l\"{o}p, A. Z. Kiss, C. Rolfs, HP.
Trautvetter, U. Greife, M. Junker, S. Goriely, M. Arnould, M. Rayet,
T. Rauscher, and H. Oberhummer, Astron. Astrophys. \textbf{333},
1112 (1998).

\bibitem{nemeth94}
Zs. N\'{e}meth, F. K\"{a}ppeler, C. Theis, T. Belgya,
and S. W. Yates, Astrophys. J. \textbf{426}, 357-365 (1994).

\bibitem{Dillmann08}
I Dillmann, T. Rauscher, M. Heil, F. K\"{a}ppeler, W. Rapp, and F.
K-. Thielemann, J. Phys. G: Nucl. Part. Phys. \textbf{35}, 0104029
(2008).

\bibitem{Babishov06}
E. M. Babishov and I. V. Kopytin, Astron. Rep.
\textbf{50}, 569-578 (2006).

\bibitem{Rayet95}
M. Rayet, M. Arnould, M. Hashimoto, et al., Astron. Astrophys.
\textbf{298}, 517 (1990).

\bibitem{Rausher02}
T. Rauscher, A. Heger, R. D. Hoffman, and S. E. Woosley,
Astrophys. J. \textbf{576}, 323 (2002).

\bibitem{Ozkan02}
N. \"{O}zkan, A. St. J. Murphy, R. N. Boyd, A. L. Cole, M. Famiano,
R. T. G\"{u}ray, M. Howard, L. Sahin, J. J. Zack, R. deHaan, J.
G\"{o}rres, M. C. Wiescher, M. S. Islam, and T. Rauscher, Nucl.
Phys. \textbf{A710}, 469 (2002).

\bibitem{nudat}
http://www.nndc.bnl.gov/nudat2/.

\bibitem{Blachot02}
J. Blachot, Nucear Data Sheet, \textbf{95}, 679 (2002).


\bibitem{Rolfs87}
C.E. Rolfs and W.S. Rodney, {\it Cauldrons in the Cosmos} (The University of Chicago Press, Chicago, 1988).

\bibitem{Debertin}
K. Debertin and R.G. Helmer, {\it Gamma-And X-ray Spectrometry With
Semiconductor Detectors} (North-Holland, Amsterdam, 1989).

\bibitem{Iliad07}
C. Iliadis, \textit{Nuclear Physics of Stars} (Wiley-VCH, Weinheim, 2007).

\bibitem{SRIM}
J.P. Biersack and J.F. Ziegler, SRIM code Version SRIM-2008.04.

\bibitem{nonsmokerweb}
T. Rauscher, code NON-SMOKER$^\mathrm{WEB}$, http://nucastro.org/websmoker.html.

\bibitem{RT98}
T. Rauscher and F.-K. Thielemann, in \textit{Stellar Evolution, Stellar Explosions And Galactic Chemical Evolution}, ed.\ A. Mezzacappa (IOP, Bristol 1998), p.\ 519.

\bibitem{Raus00}
T. Rauscher and F.-K. Thielemann, At.\ Data Nucl.\ Data Tables \textbf{75}, 1 (2000).

\bibitem{Raus01}
T. Rauscher and F. K. Thielemann, At. Data Nucl. Data Tables
\textbf{79}, 47 (2001).

\bibitem{McFadden}
McFadden and G. R. Satchler, Nucl. Phys. \textbf{84}, 177 (1966).

\bibitem{Frohlich}
C. Fr\"{o}hlich, Master thesis (University of Basel, Switzerland,
2002).

\bibitem{Rauscher03}
T. Rauscher, Nucl.\ Phys.\ \textbf{A719}, 73c (2003); erratum: T. Rauscher, Nucl.\ Phys.\ \textbf{A725}, 295 (2003).

\bibitem{Avrigeanu}
M. Avrigeanu, W. von Oertzen, A.J. M. Plompen, and V. Avrigeanu,
Nucl. Phys. \textbf{A723} 104 (2003).

\bibitem{Koehler04}
P. E. Koehler, Yu. M. Gledenov, T. Rauscher, and C. Fr\"{o}hlich,
Phys. Rev. C \textbf{69}, 015803 (2004).

\bibitem{Descouvemont06} P. Descouvemont, T. Rauscher,
Nucl.\ Phys.\ \textbf{A777}, 137 (2006).

\end{thebibliography}

\end{document}